\documentclass[11pt,cleveref, autoref]{article}

\usepackage{dblfloatfix,empheq,thm-restate,graphicx,verbatim,amsmath,amsfonts,amssymb,amsthm,alltt,algorithm,algorithmic,epsfig,blindtext,multicol,fullpage,url,authblk,titlesec,setspace,stmaryrd}
\usepackage[margin=1in]{geometry}
\usepackage[sortcites=true,maxnames=1000]{biblatex}
\usepackage[margin=0pt,font={small},labelfont={bf},labelsep=colon,format=plain]{caption}
\usepackage[left]{lineno}
\usepackage[mathscr]{eucal}
\usepackage[hidelinks]{hyperref}

\newtheorem{theorem}{Theorem}

\newcounter{remarks}
\newtheorem{remark}[remarks]{Remark}
\newcounter{dugmaot}
\newtheorem{example}[dugmaot]{Example}

\def\boldhead#1:{\par\vskip 7pt\noindent{\bf #1:}\hskip 10pt}
\def\ithead#1:{\par\vskip 7pt\noindent{\it #1:}\hskip 10pt}

\def\inline#1:{\par\vskip 7pt\noindent{\bf #1:}\hskip 10pt}
\def\midinline#1:{\par\noindent{\bf #1:}\hskip 10pt}
\def\dnsinline#1:{\par\vskip -7pt\noindent{\bf #1:}\hskip 10pt}
\def\ddnsinline#1:{\newline{\bf #1:}\hskip 10pt}
\def\largeinline#1:{\par\vskip 7pt\noindent{\large\bf #1:}\hskip 10pt}
\long\def\comment #1\commentend{}
\long\def\commhide #1\commhideend{}
\long\def\commfull #1\commend{#1}
\long\def\commabs #1\commenda{}
\long\def\commtim #1\commendt{#1}
\long\def\commb #1\commbend{}
\long\def\commedit #1\commeditend{} % Editing comments, marked also by $>>>$

\long\def\commB #1\commBend{}       % Omit in 1996 (both TR and Siena)
\long\def\commex #1\commexend{}     % LN home exercise (hide solutions)

\long\def\commsiena #1\commsienaend{}  % omit in Siena, show in TR

\long\def\commBI #1\commBIend{}  % omit in Bar-Ilan

\long\def\CProof #1\CQED{}

\def\blackslug{\hbox{\hskip 1pt \vrule width 4pt height 8pt
    depth 1.5pt \hskip 1pt}}
\def\QED{\quad\blackslug\lower 8.5pt\null\par}
\long\def\PPP#1{\noindent{\bf Proof:}{ #1}{\quad\blackslug\lower 8.5pt\null}}

\long\def\denspar #1\densend
{#1}
\newcommand{\set}[1]{\left\{ #1 \right\}}

\newif\ifnotesw\noteswtrue% T to show box & marginal notes; F supresses.
   {\ifnotesw\marginpar[\hfill\(\top\)]{\(\top\)}\fi}%
   {\ifnotesw\marginpar[\hfill\(\bot\)]{\(\bot\)}\fi}

\newcommand{\mnote}[1]%
    {\ifnotesw\marginpar%
        [{\scriptsize\it\begin{minipage}[t]{\marginparwidth}
        \raggedleft#1%
                        \end{minipage}}]%
        {\scriptsize\it\begin{minipage}[t]{\marginparwidth}
        \raggedright#1%
                        \end{minipage}}%
    \fi}

\def\MathF{\hbox{\rm I\kern-2pt F}}
\def\MathP{\hbox{\rm I\kern-2pt P}}
\def\MathR{\hbox{\rm I\kern-2pt R}}
\def\MathZ{\hbox{\sf Z\kern-4pt Z}}
\def\MathN{\hbox{\rm I\kern-2pt I\kern-3.1pt N}}
\def\MathC{\hbox{\rm \kern0.7pt\raise0.8pt\hbox{\footnotesize I}
\kern-4.2pt C}}
\def\MathQ{\hbox{\rm I\kern-6pt Q}}

\newsavebox{\ttop}\newsavebox{\bbot}

\def\nin{{~\not \in~}}

\def\titlename{}
\def\authorsnames{}

\def\nin{{~\not \in~}}

\newcommand{\qoute}[1]{``#1''}
\newcommand*{\settitlename}[1]{\title{#1}\def\titlename{#1}}
\newcommand*{\addauthor}[2]{\author[#2]{#1} 
\IfNoValueTF{\authorsnames}{\def\authorsnames{#1}}{\expandafter\def\expandafter\authorsnames\expandafter{\authorsnames, #1}}}
\newcommand{\keywords}[1]{\def\keywordnames{#1}}
\newcommand{\printkeywords}{%
    \par
    \vspace{0.5em}
    \noindent
    \textbf{Keywords: }
    \keywordnames
    \par
}

\newtheorem{question}{Question}
\theoremstyle{definition}
\newtheorem{definition}{Definition}
\newcommand {\secref} [1] {\hyperref[sec:#1]{Section \ref{sec:#1}}}
\newcommand {\subsecref} [1] {\hyperref[subsec:#1]{Subsection \ref{subsec:#1}}}
\newcommand {\thmref} [1] {\hyperref[thm:#1]{Theorem \ref{thm:#1}}}
\newcommand {\lemref} [1] {\hyperref[lem:#1]{Lemma \ref{lem:#1}}}
\newcommand {\exmref} [1] {\hyperref[exm:#1]{Example \ref{exm:#1}}}
\newcommand {\figref} [1] {\hyperref[fig:#1]{Figure \ref{fig:#1}}}
\newcommand {\defref} [1] {\hyperref[def:#1]{Definition \ref{def:#1}}}
\newcommand {\queref} [1] {\hyperref[que:#1]{Question \ref{que:#1}}}
\newcommand {\labelsec} [1] { \label{sec:#1}}
\newcommand {\labelsubsec} [1] { \label{subsec:#1}}
\newcommand {\labelthm} [1] { \label{thm:#1}}

\newcommand {\labelexm} [1] { \label{exm:#1}}
\newcommand {\labelfig} [1] { \label{fig:#1}}
\newcommand {\labeldef} [1] { \label{def:#1}}
\newcommand {\labelque} [1] { \label{que:#1}}

\newcommand{\algref}[1]{\hyperref[alg:#1]{Algorithm~\ref*{alg:#1}}}

\newcommand{\cncref}[1]{~\protect\hyperref[cnc:#1]{Conclusion~\ref*{cnc:#1}}}
\newcommand{\corref}[1]{~\protect\hyperref[cor:#1]{Corollary~\ref*{cor:#1}}}

\newcommand{\obvref}[1]{\hyperref[obv:#1]{Observation~\ref*{obv:#1}}}
\newcommand{\clmref}[1]{\hyperref[clm:#1]{Claim~\ref*{clm:#1}}}
\newcommand{\stpref}[3]{\hyperref[alg:#1_#2]{\textit{#3}}}
\NewDocumentCommand\mprod{mgg}{\displaystyle \prod\IfNoValueTF{#2}{}{_{#2}}\IfNoValueTF{#3}{}{^{#3}}{#1}}
\NewDocumentCommand\msum{mgg}{\displaystyle \sum\IfNoValueTF{#2}{}{_{#2}}\IfNoValueTF{#3}{}{^{#3}}{#1}}
\NewDocumentCommand\mcup{mgg}{\displaystyle \bigcup\IfNoValueTF{#2}{}{_{#2}}\IfNoValueTF{#3}{}{^{#3}}{#1}}
\NewDocumentCommand\mcap{mgg}{\displaystyle \bigcap\IfNoValueTF{#2}{}{_{#2}}\IfNoValueTF{#3}{}{^{#3}}{#1}}

\settitlename{Dependency Preservation May Prevent Stopping at $2NF$}

\addauthor{Amir Sapir}{1,$*$}
\addauthor{Ariel Sapir}{2,$\S$}

\affil[1]{Department of \href{https://www.sapir.ac.il/en/ba/computer_science}{Computer Science}, Sapir Academic College, Sha’ar HaNegev, Israel.}
\affil[2]{Department of \href{http://cs.biu.ac.il/}{Computer Science}, Bar-Ilan University, Ramat-Gan, Israel.}
\affil[$*$]{amirsa@post.bgu.ac.il}
\affil[$\S$]{sapirar@biu.ac.il}

\date{\today}


\makeatletter
\def\@maketitle{%
  \newpage
  \null
  \vskip 2em%
  \begin{center}%
  \let \footnote \thanks
    {\huge \textsf{\textbf{\@title}} \par}%
    \vskip 1.5em%
    {\large
      \lineskip .5em%
      \begin{tabular}[t]{c}%
        \@author
      \end{tabular}\par}%
    \vskip 1em
    {\large \@date}
  \end{center}
  \par
  \vskip 1.5em}
\makeatother

\makeatletter
\renewenvironment{abstract}{
    \begin{center}%
    {\bfseries\sffamily \Large\abstractname\vspace{\z@}}
      \end{center}%
      \quotation
    }
    \endquotation
\makeatother

\titleformat*{\section}{\Large\bfseries\sffamily}
\titleformat*{\subsection}{\large\bfseries\sffamily}

\begin{document}
\fontfamily{cmss} \selectfont

% Removing spacing regarding \left and \right brackets
\let\originalleft\left
\let\originalright\right
\renewcommand{\left}{\mathopen{}\mathclose\bgroup\originalleft}
\renewcommand{\right}{\aftergroup\egroup\originalright}

% Reducing space after figures (and tables)
\setlength{\textfloatsep}{8pt}

% Enumerating the lines
% \linenumbers

% Do not print page number for first page
\pagenumbering{gobble}

%%%%%%%%%%%%%%%%%%%%%%%%%%%%%%%%%%%%%%%%%%%%%%%%
%%%%%%%%%%%%%%%%%%% Abstract %%%%%%%%%%%%%%%%%%%
%%%%%%%%%%%%%%%%%%%%%%%%%%%%%%%%%%%%%%%%%%%%%%%%
% Printing the title
\maketitle

% Abstract
\begin{abstract}
\small 
Traditionally, it was accepted that a relational database can be normalized step-by-step, from a set of un-normalized tables to tables in $1NF$, then to $2NF$, then to $3NF$, then (possibly) to $BCNF$. In particular, the rule applied to a table in $1NF$ in order to transform it to a set of tables in $2NF$ seems to be too straightforward to pose any difficulty.

While it is a common belief to consider, theoretically, a database to be 'better' the higher it is normalized, practical usage advocates that this may not always be true -- in some cases a database performance may increase if left in a lower normal form, without sacrificing any of the advantages of a higher normal form.

It was taken for granted that a normalization process can be stopped after reaching any normal form, without proceeding to the next higher one. However, we show that, depending on the set of functional dependencies, it may be impossible to be `precisely' (in a manner to be explained in the sequel) in $2NF$. One must, in these cases, either perform the normalization from $1NF$ to $3NF$ as an indecomposable move, or settle for a normalization between $2NF$ and $3NF$. 

For a clear presentation and a concise characterization of the phenomena, we model the functional dependencies as a (particular type of) directed hyper-graph -- an FD-graph. The minimal setup to exhibit the phenomena requires a single composite key, and two partially overlapping chains of transitive dependencies. This can be sketched as a specific sub-hyper-graph. Thus, an FD-graph containing that sub-hyper-graph indicates that its corresponding database cannot be precisely in $2NF$.   
\end {abstract}

% Printing the keywords
\printkeywords

% Moving to a new page
\newpage

% Setting a wider stretch, after the abstract
\setstretch{1.3}

% Start page numbering at the introduction
\pagenumbering{arabic}

%%%%%%%%%%%%%%%%%%%%%%%%%%%%%%%%%%%%%%%%%%%%%%%%
%%%%%%%%%%%%%%%%% Introduction %%%%%%%%%%%%%%%%% 
%%%%%%%%%%%%%%%%%%%%%%%%%%%%%%%%%%%%%%%%%%%%%%%%
\section{Introduction}~\labelsec{introduction}
The theory of relational databases has been established about 5 decades ago, in a fundamental paper \cite{Codd70}. One of its pillars is the notion of {\it normal forms}, which has been further developed in  \cite{Kent83}, \cite{BeFaHo1977}, \cite{Fagin79} as well as in several other papers (cf. \cite{ViSr93} for a mathematical discussion of 3NF vs. BCNF, and \cite{DeHeLiMu1992} for representation of normal forms as semi-lattices), and put on strong mathematical foundations. This topic constitutes an important part of any recognized textbook on databases, such as \cite{SiKoSu2006}, \cite{MaNa2010}, \cite{GmhUjdWj2013}.

The tendency to reach 'ultimate' normalization led to ideas as to a definition of an alternative (to BCNF) normal form \cite{Za82}, suggestion how to present, compare and refine various decompositions \cite{MaRa98}, and a determination of a necessary and sufficient condition of a relation in BCNF to be also in 4NF \cite{Mok97}. Comparison of BCNF to object-based NF is discussed in \cite{Biskup89}.

Algorithmically, in \cite{Osborn79} an exponential-time algorithm for testing the existence of a BCNF is presented. An algorithm for automatic normalization is presented in \cite{BaNaBa2008}.

The mathematical aspect has been studied too. The issue of redundancy vs. dependency preservation is discussed
in \cite{KoLi2006}. A recent paper \cite{KohLin2018} presents a new class of functional dependencies for which it is always possible to reach elimination of data redundancy.

Another direction tackles the problem of inconsistent databases, where integrity constraints imposed by functional dependencies are violated \cite{LiKiSr2018}. In that paper, the authors study the complexity of computing optimal repairs of two types of a database having inconsistencies.

There exists an algorithm which enables us to reach $3NF$ without moving through the sequence of normal forms (cf. \cite[pp. 289--292]{SiKoSu2006}). Yet, originally the subject has been introduced as a sequential process, in which a table which is in a certain normal form is transformed to the next one by applying an additional specific demand to it, so that the tables created by the decomposition (application of the process) adhere to the specific stronger demand, in addition to the previous ones.

Classical presentations (e.g. \cite{Kent83,MaNa2010}) describe $1NF$, $2NF$ and $3NF$ as successive stages of normalization.
It was boldly stated in textbooks that any database can be presented in any of the first three normal forms.
This is also inherent in the sequential process, it was a common belief that
\qoute{A table in $2NF$ can be transformed to a set of tables in $3NF$ provided that ...} and that \qoute{Any database can be presented in $3NF$} encapsulate together the assumption that one could, if one wanted, keep some of the tables in $2NF$, without applying the additional decomposition rule associated with $3NF$. This raises the following question:

\begin{question}
\labelque{precise2NF}
Given an initial table in $1NF$ and a set of
functional dependencies, does there always exist a database in $2NF$ that can be obtained using only decomposition steps based on partial dependencies on candidate keys?
\end{question}

We answer \queref{precise2NF} negatively. We formalize the permitted operation as a $2NF$ decomposition step and call a database obtainable through such steps \emph{precisely in $2NF$}. We show that, for some sets of functional dependencies, applying only the $2NF$ decomposition rule cannot produce a proper $2NF$ database. Nevertheless, the initial table may still admit a proper decomposition directly into $3NF$. We then give a sufficient condition, expressed in terms of the functional dependencies, under which a database  cannot be precisely in $2NF$.

In \secref{preliminaries}, the notations, definitions and setup required for the rest of the paper are presented. Two types of problems of $2NF$ are described in \secref{methodical}; these are problems of a technical nature. The main problem, that of situations for which it is impossible to normalize precisely to $2NF$, is stated in \secref{inability2NF}. This shows that, mathematically speaking, $2NF$ is not a sound normal form, concluding the theme of the paper.
% Korth2014p

%%%%%%%%%%%%%%%%%%%%%%%%%%%%%%%%%%%%%%%%%%%%%%%%
%%%%%%%%%%%%%%%%% Prelimineries %%%%%%%%%%%%%%%%% 
%%%%%%%%%%%%%%%%%%%%%%%%%%%%%%%%%%%%%%%%%%%%%%%%
\section{Preliminaries}~\labelsec{preliminaries}
The notations we use are as customary in the field. Afterwards we present a required definition and several statements pertaining to proper decompositions. Then we present a simplified setup, to enable focus on the important points of the paper.

\subsection{Notations}~\labelsubsec{notations}
Let $R_{i}, i=1,2,\ldots, m$ indicate tables, $r(R_{i}) \equiv r_{i}$ the set of tuples of $R_{i}$, $\mathcal{R} = \{ R_{1}, \ldots, R_{m}\}$ a database. An $A_{i}$ denotes an attribute and $\alpha, \beta, \gamma$ non-empty sets of attributes. The set of all attributes $\Omega = A_{i}, i=1,2,\ldots, n$ will be referred to as a single initial table $R_{\Omega}$, in $1NF$, from which the normalization process starts. 
$\alpha \rightarrow \beta$ stands for a functional dependency (henceforth {\it fd}), and $\alpha \rightarrow R_{i}$ designates that $\alpha$ is a (super) key of $R_{i}$. The set of fd's is $F$ and its closure is $F^{+}$. In listing the attributes of a table, an underline below an attribute (a set of attributes) indicates that it is a candidate key -- a key with no unnecessary attributes -- of the table.

\begin{example}~\labelexm{exmp1}
Let $\Omega = \{A_1, A_2, A_3\}$ with functional dependencies $F = \{A_{1} \rightarrow A_{2}, A_{1} \rightarrow A_{3}\}$. Then
$\mathcal{R}_{a} = \{R_{1}\}$ where $R_{1} = \{\underline{A_1}, A_2, A_3\}$ is a decomposition. It is not the only one -- $\mathcal{R}_{b} = \{R_{1}, R_{2}\}$ with $R_{1} = \{\underline{A_1}, A_2\}, R_{2} = \{\underline{A_1}, A_3\}$ being another decomposition.
\end{example}

\subsection{Definitions}~\labelsubsec{definitions}
We mention the basic definitions of normal forms. All are given with respect to a given set $\Omega$ of all attributes and to a given set $F$ of functional dependencies. The definitions are taken from \cite{SiKoSu2006}.

% Atomic attribute
\begin{definition}~\labeldef{atomic}
{\normalfont \fontfamily{cmss} \selectfont 
An attribute is {\it atomic} if it assumes a value from a domain in which all the values are indivisible (e.g. integers, strings,...). }
\end{definition}

% 1NF
\begin{definition}~\labeldef{1nf}
{\normalfont \fontfamily{cmss} \selectfont A table $R_{i}$ is in $1NF$ if each of its attributes is atomic. A database $R$ is in $1NF$ if all of its tables are in $1NF$.}
\end{definition}

% prime
\begin{definition}~\labeldef{prime}
{\normalfont \fontfamily{cmss} \selectfont Given a table $R_{i}$, an attribute $A\in R_{i}$ is {\it prime} if it is contained in one $R_{i}$'s candidate keys. Otherwise, $A$ is   {\it non-prime}.}
\end{definition}

% (full, partial) functional dependency 
\begin{definition}~\labeldef{fullDependency}
{\normalfont \fontfamily{cmss} \selectfont A functional dependency $\alpha\rightarrow\beta$ is
{\it partial} if
$\alpha^{\prime}\rightarrow\beta$ holds for some
$\alpha^{\prime}\subsetneq\alpha$. Otherwise, it is {\it full}.}
\end{definition}

% dependency preservation
\begin{definition}~\labeldef{dependency}
{\normalfont \fontfamily{cmss} \selectfont Given a set $F$ of functional dependencies, its closure $F^{+}$ is the set of all functional dependencies which can be deduced from $F$.  Given a table $R_{i}$, the projection $F_{i}$ of $F^{+}$ on $R_{i}$ is:
$$F_{i} = \{\alpha \rightarrow \beta \in F^{+} : \alpha,\beta \subseteq R_{i} \}.$$
Then a decomposition of $\Omega$ into a database $\mathcal{R}$ is {\it dependency preserving} when: 
$$\left( \bigcup_{R_{i} \in R} F_{i} \right)^{+} = F^{+}.$$}
\end{definition}

% lossless decomposition
\begin{definition}~\labeldef{data}
{\normalfont \fontfamily{cmss} \selectfont A decomposition of a database $\mathcal{R}=\left\{ R_1 ,R_2, ..., R_m \right\}$ into tables $R_{1,1},...,R_{1,k_1},$ $...,R_{m,1},...,R_{m,k_m}$  is  {\it lossless} if, for each $1\leq i \leq m$, a natural join of the $k_i$ tables $R_{i,1},R_{i,2},...,R_{i,k_i}$ yields back the original table $R_{i}$.}
\end{definition}

% def of 2NF
\begin{definition}~\labeldef{2nf}
{\normalfont \fontfamily{cmss} \selectfont A table $R_{i}$ is in $2NF$ if it is in $1NF$ and there is no non-prime attribute that depends on a proper part of a candidate key. A database $\mathcal{R}$ is in $2NF$ if all of its tables are in $2NF$ and $\mathcal{R}$ is both lossless and dependency preserving.}
\end{definition}

% transitive dependency
\begin{definition}~\labeldef{trans}
{\normalfont \fontfamily{cmss} \selectfont An attribute $A$ is {\it transitively dependent} on $\alpha$ if 
there exists a set of attributes $\beta\nsubseteq \alpha$ such that:
$
\alpha\rightarrow\beta,
\qquad
\beta\rightarrow A,
\qquad
\beta\nrightarrow\alpha
$
and $A\notin\alpha ,\beta$.}
\end{definition}

% 3NF
\begin{definition}~\labeldef{3nf}
{\normalfont \fontfamily{cmss} \selectfont A table $R_{i}$ is in $3NF$ if it is in $2NF$ and, for any transitive dependence $\alpha \rightarrow \beta \rightarrow A$ such that $\alpha, \beta, A$ belong to the same table $R_{i}$, the attribute $A$ is prime. A database $\mathcal{R}$ is in $3NF$ if all of its tables are in $3NF$ and $\mathcal{R}$ is both lossless and dependency preserving.}
\end{definition}

To this end, we need the following, specialized definitions.

% 2NF Step
\begin{definition}
\labeldef{2nfStep}
{\normalfont \fontfamily{cmss} \selectfont
Let $R_{i}$ be a table.  A $2NF$ \emph{decomposition step} selects a full functional dependency $\alpha\rightarrow\beta\in F_{i}$ such that $\alpha$ is a proper subset of a candidate key of $R_{i}$ and $\beta$ is a nonempty set of non-prime attributes. The step replaces $R_{i}$ by the  decomposition $\set{R_{i,0},R_{i,1}}$, where $R_{i,0}=R_{i}\setminus\beta$ and $R_{i,1}=\underline{\alpha}\cup\beta$.}
\end{definition}

% Exact 2NF
\begin{definition}~\labeldef{p2nf}
{\normalfont \fontfamily{cmss} \selectfont 
A database $\mathcal{R}$ is {\it precisely in $2NF$} if it is in $2NF$ and there exists a finite sequence of databases $\mathcal{R}^{(0)},\mathcal{R}^{(1)},\ldots,\mathcal{R}^{(q)}$ such that $\mathcal{R}^{(0)}=\Omega$, $\mathcal{R}^{(q)}=\mathcal{R}$ and, for every $0\leq j<q$, the database $\mathcal{R}^{(j+1)}$ is obtained from $\mathcal{R}^{(j)}$ by applying a $2NF$ decomposition step to one of its tables.
%Previous: A database $R$ is {\it precisely in $2NF$} if and only if it holds that: it is in $2NF$ and, starting from $R_\Omega$, the only decomposition steps taken were based on the criterion for $2NF$, with no further normalization rules.
}
\end{definition}

\begin{example}~\labelexm{exmp2}
Let $\Omega = \set{\underline{A_1, A_2}, A_3, A_4, A_5, A_6, A_7}$ be a set of attributes together with the set of fds 
$
  F= \set{ A_1 A_2 \rightarrow A_7, A_{1} \rightarrow A_{3}, A_{2} \rightarrow  A_{4},   A_{4} \rightarrow A_{5}, A_{5} \rightarrow A_{6}}
$. 
%$$F = \{A_1 A_2 \rightarrow A_7, A_{1} \rightarrow A_{3}, A_{2} \rightarrow A_{4}, A_{4} \rightarrow A_{5}, A_{5} \rightarrow A_{6}\}.$$    Consider the following decompositions:
Consider
\begin{itemize}
\item $\mathcal{R}_{a}=\{R_{1}, R_{2}, R_{3}, R_{4}, R_{5}\}$ with
  \begin{eqnarray*}
  R_{1} = \{\underline{A_1, A_2}, A_7\}, R_{2} = \{\underline{A_1}, A_3\}, \\
  R_{3} = \{\underline{A_2}, A_4\}, R_{4} = \{\underline{A_4}, A_5\}, R_{5} = \{\underline{A_5}, A_6\}.
  \end{eqnarray*}
%$\mbox{\scriptsize $R_{1} = \{\underline{A_1, A_2}, A_7\}, R_{2} = \{\underline{A_1}, A_3\}, R_{3} = \{\underline{A_2}, A_4\},\\ R_{4} = \{\underline{A_4}, A_5\}, R_{5} = \{\underline{A_5}, A_6\}.$}$
\item $\mathcal{R}_{b}=\{R_{1}, R_{2}, R_{3}, R_{4}\}$ with
  \begin{eqnarray*}
  R_{1} = \{\underline{A_1, A_2}, A_7\}, R_{2} = \{\underline{A_1}, A_3\}, \\
  R_{3} = \{\underline{A_2}, A_4\}, R_{4} = \{\underline{A_4}, A_5, A_6\}.
  \end{eqnarray*}
\item $\mathcal{R}_{c}=\{R_{1}, R_{2}, R_{3}\}$ with
  \begin{eqnarray*}
  \mbox{$R_{1} \!=\! \{\underline{A_1, A_2}, A_7\}, R_{2} \!=\! \{\underline{A_1}, A_3\}, R_{3} \!=\! \{\underline{A_2}, A_4, A_5, A_6\}$}.
  \end{eqnarray*}
\end{itemize}
Then $\mathcal{R}_{a}$ is in $3NF$ (so also in $2NF$). However, several decompositions due to transitivity took place in $\mathcal{R}_{a}$. Hence, $\mathcal{R}_{a}$ is not precisely in $2NF$. As for $\mathcal{R}_{b}$ and $\mathcal{R}_{c}$, they are in $2NF$ too. Yet, there are some transitive dependencies left in tables, so neither is in $3NF$. In $\mathcal{R}_{b}$, the tables $R_{3}$ and $R_{4}$ are a consequence of applying the rule for $3NF$, thus it is not precisely in $2NF$. However, in $\mathcal{R}_{c}$, no decomposition based on transitivity took place, so it is precisely in $2NF$.
\end{example}

\begin{remark}~\label{rem1}
{\normalfont \fontfamily{cmss} \selectfont A set of functional dependencies may contain no transitive dependencies. In this case, a decomposition to $2NF$ will be (in a trivial manner) also a decomposition to $3NF$. In such a scenario, this decomposition is still considered precisely in $2NF$, as there is no option to stop in $2NF$ without being in $3NF$ as well. In the following example, we will take a look at such a scenario.}
\end{remark}

\begin{example}~\label{exmp3}
Let $\Omega = \{\underline{A_1, A_2}, A_3, A_4, A_5\}$ with $$F = \{A_1 A_2 \rightarrow A_5, A_{1} \rightarrow A_{3}, A_{2} \rightarrow A_{4}\}.$$ The $2NF$ decomposition $\mathcal{R}_{a}=\{R_{1}, R_{2}, R_{3}\},$ with
 $$R_{1} = \{\underline{A_1, A_2}, A_5\}, R_{2} = \{\underline{A_1}, A_3\}, R_{3} = \{\underline{A_2}, A_4\}$$
%\begin{eqnarray*}
%  \mbox{\scriptsize $R_{1} = \{\underline{A_1, A_2}, A_5\}, R_{2} = \{\underline{A_1}, A_3\}, R_{3} = \{\underline{A_2}, A_4\}$}.
%\end{eqnarray*}
is, at the same time, also in $3NF$ (since no transitivity is left) and also precisely in $2NF$ (in the trivial manner -- no action based on transitivity took place).
\end{example}
For the main theme of the paper, at \secref{inability2NF}, the following definitions are required:
% chain of transitive dependencies  
\begin{definition}~\labeldef{transChains}
{\normalfont \fontfamily{cmss} \selectfont A {\it chain of transitive dependencies} is a set of functional dependencies $\{\alpha_{i} \rightarrow \alpha_{i+1}\}$ for which each functional dependency is full. This, in turn, implies that $\alpha_{i+1}\not\subset\alpha_{i}$. }  
\end{definition}

% maximal chain of transitive dependencies  
\begin{definition}~\labeldef{transChainsMax}
{\normalfont \fontfamily{cmss} \selectfont A chain of of transitive dependencies will be called a {\it maximal chain of transitive dependencies} if there is no other chain of transitive dependencies containing it.}
\end{definition}

%\begin{definition}~\labeldef{transChainsDifferent}
%{\normalfont \fontfamily{cmss} \selectfont A chain $\left\{\alpha_{i} \rightarrow \alpha_{i+1}\right\}$ of transitive dependencies will be called a {\it divisor} of another chain $\left\{\beta_{i} \rightarrow \beta_{i+1}\right\}$ if there exists a $\delta \subset \Omega$ for which $\left\{ \alpha_i, \delta \right\} = \left\{ \beta_i \right\}$. Similarly, a pair of chains will be called {\it distinct} if they have no common divisor, and a chain will be called {\it prime} if it has no divisor.}
%\end{definition}
\begin{definition}~\labeldef{transChainsOverlap}
{\normalfont \fontfamily{cmss} \selectfont
 Let $k \geq 2$ and $l\geq 1$ (or vice versa). A pair of {\it partially overlapping chains of transitive dependencies} is composed of two maximal chains of transitive dependencies $\left\{\alpha_{i} \rightarrow \alpha_{i+1}\right\}, 1 \le i \le k-1$ and $\left\{\beta_{j} \rightarrow \beta_{j+1}\right\}, 1 \le j \le l-1$ having a `meeting point' $k^{*}\leq k,l^{*}\leq l$ in which $\{\alpha_{k^{*}}\beta_{l^{*}} \rightarrow \gamma\}$. Further, for any $i$ and $j$ it holds that $\alpha_{i}\beta_{j} \rightarrow \gamma \notin F^{+}$. Additionally,
 %all $\alpha_1,\ldots,\alpha_{k}, \beta_1,\ldots,\beta_{l}, \gamma$ are disjoint and,
 for any $\delta \subset \alpha_{k^{*}}\beta_{l^{*}}$, it holds that $\delta \rightarrow \gamma \notin F^{+}$.}
\end{definition}

\begin{figure}[H]~\labelfig{chainsPic}
\begin{center}
\includegraphics[width=\columnwidth*9/10*9/10]{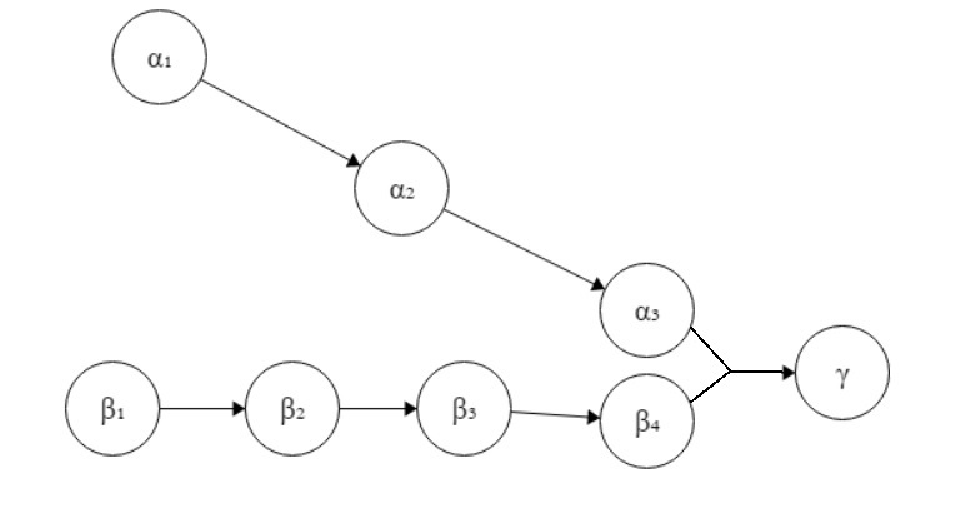}
\caption{A graph depicting an example for partially overlapping chains of transitive dependencies. Here $k^{*}=k=3$, $l^{*}=l=4$.}
\end{center}
\end{figure}

An example of a minimal-size system of partially overlapping chains of transitive dependencies follows.
\begin{example}~\labelexm{exmin}
{\rm \fontfamily{cmss} \selectfont Let $\Omega = \{\underline{A_1, A_2}, A_3, A_4\}$ with $$F = \{A_{1} \rightarrow A_{3}, \, A_{2}A_{3} \rightarrow A_{4}\},$$
where $\alpha_1 \!=\! \{A_1\}, \alpha_2 \!=\! \{A_3\}, \beta_1 \!=\! \{A_2\}, \gamma \!=\! \{A_4\}, k \!=\! 2, l \!=\! 1$.}
\end{example}

\subsection{Setup}~\labelsubsec{setup}
Our starting point is, in any of the following scenarios, that of a set of functional dependencies $F$ and a single source table $\Omega=\{A_1, A_2, \ldots, A_n\}$ already in $1NF$, to be decomposed to a set of tables  $R_{\sigma} = \{R_{1}, R_{2}, \ldots, R_{m}\}$ as dictated by $F$.

A full coverage of the possibilities seems to require consideration of numerous cases, due to
\begin{itemize}
\item the number of candidate keys $K_{1}, K_{2}, \ldots$ and the number of attributes in each $K_{i}$ (say, $K_{1}=\{A_1,A_2,A_3\}$),
\item for each $K_{i}$, which attributes in $\Omega$ depend solely of a specific $A_j \in K_{i}$, which depend on a combination of attributes, and which depend independently on several attributes, and
\item possible overlap schemes among the keys themselves (for example, $K_{2}=\{A_3,A_4\}$ has an overlap with $K_{1}$).
\end{itemize}

In order to keep the discussion as simple as possible, the dependencies will be such that, as far as $\Omega$ and $F$ are concerned, there is a single candidate key, and that key has two attributes. The principal possibilities of dependencies and overlaps (henceforth {\it cases}) will be described in the following section.

The discussion of the various cases will not, in general, be confined by the (non-)ability of providing real tables, since the purpose is mathematical formulation. Yet, for the main theme of the paper (\secref{inability2NF}), we will demonstrate by supplying a `real-life' example with a single table and a few dependencies.

\subsection{Modelling}~\labelsubsec{model}
A natural, though less common, way of modelling the network of fds of a database is by a directed hyper-graph. In general, a hyper-graph has a base set of vertices, and a hyper-edge is a subset of them. In a directed hyper-graph, a directed arc is a pair of subsets of the base set \cite{AuLa2017}. In \cite{AuDaSa83} and \cite{Maier80}, among some others, authors adopted a somewhat restricted form of a directed hyper-graph, in which a directed arc is a pair where the left member is a subset of the base set, but the right member is a single item element of it. For the process of normalization, we favor this approach. Yet, for the purpose of this paper, we prefer the general definition, for it allows for a succinct representation of the dependencies involved. \figref{shortChainsPic} shows, as a directed hyper-graph, the network of  fds of \exmref{exmin}. 

\begin{figure}[H]
\begin{center}
\includegraphics[width=\columnwidth*7/10*8/10*9/10]{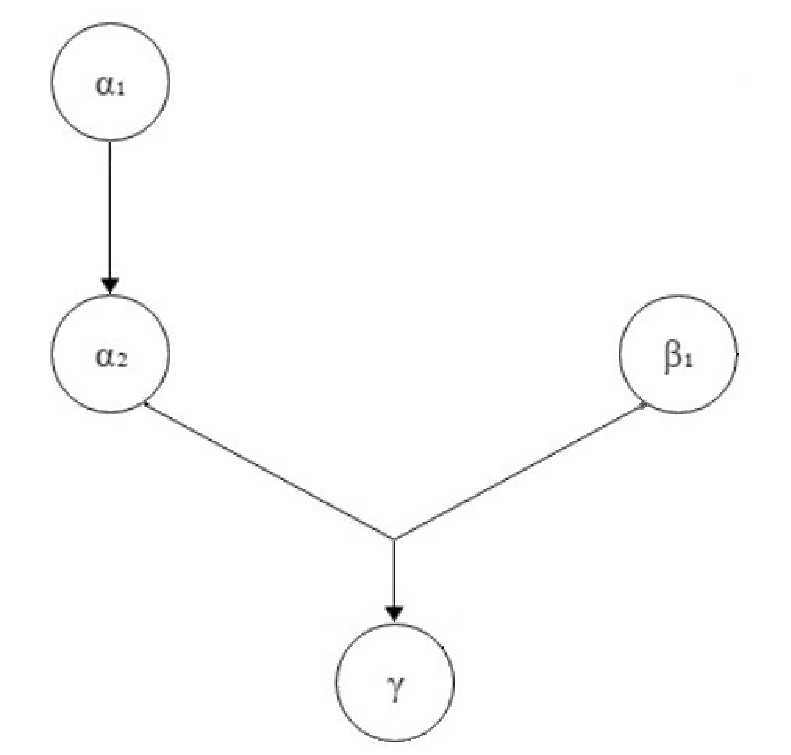}
\caption{A graph depicting \exmref{exmin} -  minimal size partially overlapping chains of transitive dependencies.}~\labelfig{shortChainsPic}
\end{center}
\end{figure}

\subsection{Motivation}~\labelsubsec{motivation}
Theoretically speaking, a relational database is 'better' if it is as normalized as possible. Practical considerations advocate that this may not always be the case, since any normalization step is carried out by splitting a table into several others and this, in turn, decreases the performance of queries (see \exmref{exmp24}). If updates are infrequent, the advantage will be neglected; if queries are $-$ time consumption due to frequent join operations will increase and we would have been better off leaving the database in between $2NF$ and higher $NF$'s. The following example, though simple by being in $2NF$ in the first place, demonstrates the issue.

\begin{example}~\labelexm{exmp24}
{\rm \fontfamily{cmss} \selectfont
Consider a table recording students ({\it sid} attribute), their departments ({\it did}) and faculties ({\it fid, fname}). The fundamental functional dependencies are
$$F = \{sid \rightarrow did, did \rightarrow fid, fid \rightarrow fname\}.$$
The $3NF$ decomposition is $\mathcal{R}_{a} = \{R_{1}, R_{2}, R_{3}\}$ where $$R_{1} = \{\underline{sid}, did\}, R_{2} = \{\underline{did}, fid\}, R_{3} = \{\underline{fid}, fname\}$$
This has the advantage of less storage in $r_{1}$ and gives the flexibility of a single update in $r_{2}$ in case a department will be moved to another faculty (or in $r_{3}$, in case the faculty name will be changed) $-$ very rare updates. 

However, since a query on $\mathcal{R}_{a}$ usually involves the natural join $(r_{1} \bowtie r_{2}) \bowtie r_{3}$ (or a sequentual pass on $r_{1}$ and corresponding searches by a $B^{+}tree$ in $r_{2}, r_{3}$), this requires much more I/O compared to $\mathcal{R}_{b} = \{R_{1}\}$ where 
$$R_{1} = \{\underline{sid}, did, fid, fname\},$$
is a $2NF$ decomposition that keeps all the data in a single table.}
\end{example}

%%%%%%%%%%%%%%%%%%%%%%%%%%%%%%%%%%%%%%%%%%%%%%%%
%%%%%%%%%%%%%%%%% To 2NF %%%%%%%%%%%%%%%%% 
%%%%%%%%%%%%%%%%%%%%%%%%%%%%%%%%%%%%%%%%%%%%%%%%
\section{Methodical normalization into 2NF}~\labelsec{methodical}
The main question the paper addresses is how to normalize a table precisely to $2NF$ in the presence of partially overlapping chains of transitive dependencies. However, our starting point is the basic, principal structures in which the setup of \subsecref{setup} usually appears. Within this scope, we consider (in \subsecref{4cases}) the relationship between the sets of attributes determined by the components of the (single, composite) key. Specifically, we ask whether they form a partition of $\Omega$: do they overlap or not, and does their union (as separate attributes!) cover all of $\Omega$. This is summed up to four principal cases, and we formulate, for each case, the procedural steps that should be carried out in order to implement the $2NF$ normalization.

In ~\subsecref{legitimate} we rule out several of them, since either preservation of data or preservation of dependencies is not maintained. This is essential for the main statement of this paper, but also important by itself since, as far as the authors know, the issue of preservation of data is kept in practice, but not stated explicitly, as seen in
\begin{example}~\labelexm{exmp6}
{\rm \fontfamily{cmss} \selectfont Let $\Omega = \{\underline{A_1, A_2}, A_3, A_4\}$ with $$F = \{A_{1} \rightarrow A_{3}, A_{2} \rightarrow A_{4}\}.$$ Strictly following the definition of $2NF$ leads to the decomposition
$$\mathcal{R}_{a}=\{R_{1}, R_{2}\},\qquad R_{1} = \{\underline{A_1}, A_3\}, R_{2} = \{\underline{A_2}, A_4\}.$$
Yet, it is obvious for the people in the field that by doing so one loses the information as to which combinations of values for $A_{1}, A_{2}$ are allowed and that (this is the essence of lossless join decomposition) the proper decomposition is
$\mathcal{R}_{b}=\{R_{1}, R_{2}, R_{3}\}$ with
%  \begin{eqnarray*}
%  \mbox{\scriptsize $R_{1} \!=\! \{\underline{A_1, A_3}, A_7\}, R_{2} \!=\! \{\underline{A_2}, A_4\}, R_{3} = \{\underline{A_1, A_2}\}$}.
%  \end{eqnarray*}
$$R_{1} = \{\underline{A_1}, A_3\}, R_{2} = \{\underline{A_2}, A_4\}, R_{3} = \{\underline{A_1, A_2}\}.$$}
\end{example}
Preservation of data of a table is achieved by decomposing the table (to, say, $R_1$ and $R_2$) in a lossless-join manner. This is characterized by that, when performing a natural join between the resulting tables, we restore precisely the original information. A simple criterion for that is by testing whether $R_{1} \cap R_{2}$ is a key in either $R_{1}$ or $R_{2}$ (cf. \cite[pp. 285--286]{SiKoSu2006}).

\subsection{The four principal cases}~\labelsubsec{4cases}
 We state the four principal cases, and give an informal argument as to why each other potential case is basically the same as one of these four.
 Common to the four cases is:
\begin{table}[H]~\label{tab1}
\begin{center}
\begin{tabular}{ll}
 $A_{1}A_{2} \rightarrow \Omega$ & \cr
 $\alpha_{1} = A_{1}^{+} \subset \Omega\, ,$   & $\alpha_{2} = A_{2}^{+} \subset \Omega \, ,$  \cr
 $A_{1} \rightarrow A_{2} \nin F^{+},$         & $A_{2} \rightarrow A_{1} \nin F^{+}$.
\end{tabular}
\end{center}
\end{table}

For brevity, we denote: $\alpha_{1}^{-} = \alpha_{1} - \{A_{1}\}$ and $\alpha_{2}^{-} = \alpha_{2} - \{A_{2}\}$. Note that, by $A_{i}^{+} \subset \Omega$, it is meant strict inclusion. Otherwise, the setup degenerates to that of a single-attribute key, which poses no problem. Lastly, 
we remark that $A_{1} \rightarrow A_{2} \nin F^{+},$ and $A_{2} \rightarrow A_{1} \nin F^{+}$ is implied by the previous demands, and is specified for clarity. The cases are:

\begin{center}
\begin{multicols}{2}
% subcase 1.1
\underline{Case 1}\\
$\alpha_{1} \cap \alpha_{2} = \varnothing, \,\, \alpha_{1} \cup \alpha_{2} = \Omega$: \\
$\begin{array}{l}
 \mathcal{R}= \{R_{1}, R_{2}\} \cr
 R_{1} = \{\underline{A_{1}}\} \cup \alpha_{1}^{-} \cr
 R_{2} = \{\underline{A_{2}}\} \cup \alpha_{2}^{-} \cr
 \cr
 \cr
\end{array}$

\columnbreak

% subcase 1.2
\underline{Case 2}\\ $\alpha_{1} \cap \alpha_{2} = \varnothing, \,\, \alpha_{1} \cup \alpha_{2} \subset \Omega$: \footnote{This is the classical case, and can be seen as an extension of Case 1.} \\
$\begin{array}{l}
 \mathcal{R}= \{R_{1}, R_{2}, R_{3}\} \cr
 R_{1} = \{\underline{A_{1}}\} \cup \alpha_{1}^{-} \cr
 R_{2} = \{\underline{A_{2}}\} \cup \alpha_{2}^{-} \cr
 R_{3} = \{\underline{A_{1},A_{2}}\} \cup (\Omega - (\alpha_{1} \cup \alpha_{2})) \cr
% A_{2} \rightarrow \alpha_{2} \subset R \cr
% A_{1} \rightarrow A_{2} \nin F^{+}, \,\,\, A_{2} \rightarrow A_{1} \nin F^{+}
 \cr
\end{array}$
\end{multicols}

\vspace{-1.5cm}

\begin{multicols}{2}
% subcase 1.3
\underline{Case 3}\\ $\alpha_{1} \cap \alpha_{2} \neq \varnothing, \,\, \alpha_{1} \cup \alpha_{2} = \Omega$.\\ 
This case is further split into $2$ subcases. In subcase (a), we include the common attributes only in one of the $R_{i}'s$; in subcase (b) -- in both of them, as follows:

\columnbreak

% subcase 1.4
\underline{Case 4}\\ $\alpha_{1} \cap \alpha_{2} \neq \varnothing, \,\, \alpha_{1} \cup \alpha_{2} \subset \Omega$.\\ This case, too, is further split into $2$ subcases. In subcase (a), we include the common attributes only in one of the $R_{i}'s$; in subcase (b) -- in both of them, as follows:

\end{multicols}

\vspace{-0.75cm}

\begin{multicols}{2}
% subcase 1.3A
\begin{itemize}
\item {3a}
$\begin{array}{l}
\mathcal{R}= \{R_{1}, R_{2}\} \cr
 R_{1} = \{\underline{A_{1}}\} \cup \alpha_{1}^{-} \cr
 R_{2} = \{\underline{A_{2}}\} \cup \left(\alpha_{2}^{-} - \alpha_{1}\right) \cr
\end{array}$
\end{itemize}
\columnbreak

% subcase 1.4A
\begin{itemize}
\item {4a}
$\begin{array}{l}
 \mathcal{R}=\{R_{1}, R_{2}, R_{3}\} \cr
 R_{1} = \{\underline{A_{1}}\} \cup \alpha_{1}^{-} \cr
 R_{2} = \{\underline{A_{2}}\} \cup \left(\alpha_{2}^{-} - \alpha_{1}\right)\cr
 R_{3} = \{\underline{A_{1},A_{2}}\} \cup \left(\Omega - \left(\alpha_{1} \cup \alpha_{2}\right)\right) \cr
\end{array}$
\end{itemize}

\end{multicols}

\begin{multicols}{2}

% subcase 1.3B
\begin{itemize}
\item {3b}
$\begin{array}{l}
 \mathcal{R}= \{R_{1}, R_{2}\} \cr
 R_{1} = \{\underline{A_{1}}\} \cup \alpha_{1}^{-} \cr
 R_{2} = \{\underline{A_{2}}\} \cup \alpha_{2}^{-} \cr
\end{array}$
\end{itemize}

\columnbreak

% subcase 1.4B
\begin{itemize}
\item {4b}
$\begin{array}{l}
\mathcal{R}= \{R_{1}, R_{2}, R_{3}\} \cr
 R_{1} = \{\underline{A_{1}}\} \cup \alpha_{1}^{-} \cr
 R_{2} = \{\underline{A_{2}}\} \cup \alpha_{2}^{-} \cr
 R_{3} = \{\underline{A_{1},A_{2}}\} \cup \left(\Omega - \left(\alpha_{1} \cup \alpha_{2}\right)\right) \cr
\end{array}$
\end{itemize}

\end{multicols}
\end{center}

\subsection{Which cases are legitimate?}~\labelsubsec{legitimate}
As stated at the beginning of the section, a proper decomposition must maintain dependency preservation and be lossless.

In this paragraph we show that only Case 2 and Case 4b are proper $2NF$ decompositions. None of the cases 1, 3a, 3b, and 4a is legitimate. Each of these will be ruled out either since it does not maintain dependencies or is not lossless join.  Then we will see that cases 1 and 3b can be 'merged' into 2 and 4b, respectively. The result will be having two main cases: case A (unification of Cases 1 and 2) and case B (unification of Cases 3b and 4b). Cases 3a and 4a will be discarded.
\\

\begin{itemize}
\item{\underline{Basic Case}} (union of Cases 1 and 2): $\alpha_{1} \cap \alpha_{2} = \varnothing$: \\
$\begin{array}{l}
 \mathcal{R}= \{R_{1}, R_{2}, R_{3}\} \cr
 R_{1} = \{\underline{A_{1}}\} \cup \alpha_{1}^{-} \cr
 R_{2} = \{\underline{A_{2}}\} \cup \alpha_{2}^{-} \cr
 R_{3} = \{\underline{A_{1},A_{2}}\} \cup (\Omega - (\alpha_{1} \cup \alpha_{2})) \cr
\end{array}$
\\For Case 1, decomposition without $R_{3}$ is lossy:
 $$R_{1} \cap R_{2} = \varnothing, r_{1 \times 2} = r_{1} \bowtie r_{2} = r_{1} \times r_{2}.$$
 It holds that $|r_{1 \times 2}| = |r_{1}| \cdot |r_{2}| \ge |r_{\Omega}|$, often with strict inequality, as discussed in ~\exmref{exmp6}. However, 'merging' Case 1 with Case 2, that is, establishing $R_{3} = \{\underline{A_{1},A_{2}}\} \cup (\Omega - (\alpha_{1} \cup \alpha_{2}))$ even when $\alpha_{1} \cup \alpha_{2} = \Omega$, results in $R_{3} = \{\underline{A_{1},A_{2}}\}$ for this case, avoiding the lossy decomposition situation. The presence of $R_3$ restricts the possible combinations of data for $A_1$ and $A_2$ to those originally present in $r_{\Omega}$:  Hence, after performing a natural join on $r_1, r_2, r_3$, we obtain
 $r_{\Omega} = r_{1} \bowtie r_{2} \bowtie r_{3}$.

\item{\underline{Case A}} (union of Cases 3a and 4a) does not preserve functional dependencies. We explain it in the following discussion, without loss of generality, for Case 4a:

Let $\beta = \alpha_{1}^{-} \cap \alpha_{2}^{-}$. Since $\beta\neq\varnothing$, there exists an attribute $A_3 \in \beta$. By definition of $\beta$, we have $A_1 \rightarrow A_3, \, A_2 \rightarrow A_3 \in F^{+}$. However, while $F_{1}$ (recall that $F_{i}$ is the set of functional dependencies of $F^{+}$ restricted to $R_{i}$) contains $A_1 \rightarrow A_3$, this is not the case for $A_2 \rightarrow A_3$: it is not included in $F_{2}$ and cannot be deduced from $\,\bigcup F_{i}$. The dependency $A_2 \rightarrow A_3$ is lost, thus making this decomposition improper.

The correct way to fix this will be by changing $R_2$ to include all the common attributes of $\beta$. This, together with the establishment of $R_{3}$ (in the same way done for the Basic Case) leads to the establishment of:
\item{\underline{Case B}} (union of Cases 3b and 4b): $\alpha_{1} \cap \alpha_{2} \neq \varnothing$: \\
$\begin{array}{l}
\mathcal{R}= \{R_{1}, R_{2}, R_{3}\} \cr
 R_{1} = \{\underline{A_{1}}\} \cup \alpha_{1}^{-} \cr
 R_{2} = \{\underline{A_{2}}\} \cup \alpha_{2}^{-} \cr
 R_{3} = \{\underline{A_{1},A_{2}}\} \cup (\Omega - (\alpha_{1} \cup \alpha_{2})) \cr
\end{array}$
\\For Case 3b, a decomposition without $R_{3}$ is lossy:
 $$R_{1} \cap R_{2} = \varnothing, r_{1 \times 2} = r_{1} \bowtie r_{2} = r_{1} \times r_{2}.$$
 It holds that $|r_{1 \times 2}| = |r_{1}| \cdot |r_{2}| \ge |r_{\Omega}|$, often with strict inequality, as discussed in ~\exmref{exmp6}. However, 'merging' Case 3b with Case 4b, that is, establishing $R_{3} = \{\underline{A_{1},A_{2}}\} \cup (\Omega - (\alpha_{1} \cup \alpha_{2}))$ even when $\alpha_{1} \cup \alpha_{2} = \Omega$, results in $R_{3} = \{\underline{A_{1},A_{2}}\}$ for this case, avoiding the lossy decomposition situation. The presence of $R_3$ restricts the possible combinations of data for $A_1$ and $A_2$ to those originally present in $r_{\Omega}$:  Hence, after performing a natural join on $r_1, r_2, r_3$, we obtain
 $r_{\Omega} = r_{1} \bowtie r_{2} \bowtie r_{3}$, precisely as done in the \underline{Basic Case} above.

\end{itemize}

%%%%%%%%%%%%%%%%%%%%%%%%%%%%%%%%%%%%%%%%%%%%%%%%
%%%%%%%%%%%%%%%%% Inability 2NF %%%%%%%%%%%%%%%%% 
%%%%%%%%%%%%%%%%%%%%%%%%%%%%%%%%%%%%%%%%%%%%%%%%
\section{Inability of being precisely in 2NF}~\labelsec{inability2NF}
The main theme of the paper is the statement of a sufficient condition for which it is impossible to normalize precisely to 2NF. We start by describing a simplified situation.
We mention, in passing, that one minimal setup to exhibit the phenomena requires a single composite key, and two partially overlapping chains of transitive dependencies, and this example is in accordance with it.
\begin{example}~\labelexm{exmp5}
\fontfamily{cmss} \selectfont A database is intended to record the eligibility of students to a fee reduction for the courses they take. A student is described by the  {\it sid} attribute, the course by the  {\it cid} attribute. The reduction {\it rd} is determined by the socio-economical status {\it st} of the student and the credit points {\it cr} of the course. Thus the attributes are $\Omega = \{\underline{sid, cid}, st, cr, rd\}$ and the fundamental functional dependencies are
$$F = \{sid \rightarrow st, cid \rightarrow cr, \{st, cr\} \rightarrow rd\}.$$
Following the definition of $2NF$ leads to the tables $\{\underline{sid}, st\}$ and $\{\underline{cid}, cr\}$. Since $\{rd\}$ has full dependence on any of the pairs $\{sid, cid\}$ and $\{st, cr\}$, there are two ways to incorporate it in a table -- one for each of the pairs, namely $\mathcal{R}_{a}=\{R_{1}, R_{2}, R_{3}\}$, where
$$R_{1} = \{\underline{sid, cid}, rd\}, R_{2} = \{\underline{sid}, st\}, R_{3} = \{\underline{cid}, cr\},$$
and $\mathcal{R}_{b}=\{R_{1}, R_{2}, R_{3}\},$ where
$$R_{1} = \{\underline{st, cr}, rd\}, R_{2} = \{\underline{sid}, st\}, R_{3} = \{\underline{cid}, cr\}.$$

Yet, none of these decompositions is appropriate. Consider $\mathcal{R}_{a}$: the dependence $\{st, cr\} \rightarrow rd$ cannot be reconstructed. In $\mathcal{R}_{b}$, the decomposition is indeed in $2NF$, but not precisely in $2NF$: $rd$ is transitively dependent on $\{sid, cid\}$, along the chain $\{sid, cid\} \rightarrow \{st, cr\} \rightarrow rd$, thus its separation to $R_{1} = \{\underline{st, cr}, rd\}$ is a $3NF$ decomposition step and should not be performed on the way to $2NF$.
\end{example}

%A sufficient 
% \subsection{Equivalent Condition}~\labelsubsec{no2NFCondition}

The following states a sufficient condition on the hyper-graph of fd's, so that the system cannot be decomposed precisely to $2NF$. It is based on the setup of \subsecref{setup}: 

\begin{theorem}~\labelthm{no2NF}
 Let $F$ be a set of functional dependencies of a database. If the directed hyper-graph describing $F$ contains a subgraph of a pair of partially overlapping chains of transitive dependence, then the database cannot be decomposed precisely into $2NF$.
\end{theorem}

%Let $\Omega$ be the single $1NF$ table of all attributes, $K = \{A_1, A_2\}$ its only candidate key, and $F$ its set of functional dependencies. If $F$ contains a pair of partially overlapping chains of transitive dependence originating at $A_{1}$ and $A_{2}$ respectively, then the system cannot be decomposed into precisely $2NF$.

%, $\alpha_{1} = A_{1}^{+}, \alpha_{2} = A_{2}^{+}, \alpha_{3} = \Omega - \{\alpha_{1} \cup \alpha_{2}\} \neq \varnothing$. If

% Let $\Omega$ be the single $1NF$ table of all attributes, $K = \{A_1, A_2\}$ its only candidate key, $\alpha_{1} = A_{1}^{+}, \alpha_{2} = A_{2}^{+}, \alpha_{3} = \Omega - \{\alpha_{1} \cup \alpha_{2}\} \neq \varnothing$. Suppose there exists $\beta$ such that $\beta \cap \alpha_{1} \neq \varnothing, \beta \cap \alpha_{2} \neq \varnothing, \beta \cap \alpha_{3} = \varnothing, A_{1}, A_{2} \nin \beta.$ Then, if there is $B \in \alpha_{3}$ such that $\beta \rightarrow B \in F^{+}$, then the system cannot be decomposed so that it will be precisely in $2NF$.

\begin{proof}
%Let $\alpha_{1}, \!\ldots, \!\alpha_{k}$ with $\{\alpha_{i} \!\!\rightarrow\!\! \alpha_{i+1}\}, 1 \!\! \le \!\! i \!\! \le \!\! k\!\!-\!\!1$ and $\beta_{1}, \ldots, \beta_{l}$ with $\{\beta_{j} \rightarrow \beta_{j+1}\}, 1 \le j \le l-1$ be the pair of {\it partially overlapping chains of transitive dependencies} with all the requirements as in
Let $\alpha_{1}, \ldots, \!\alpha_{k}$ with $\{\alpha_{i} \!\!\rightarrow\!\! \alpha_{i+1}\}, 1 \!\le\! i \!\le\! k\!-\!1$ and $\beta_{1}, \ldots, \beta_{l}$ with $\{\beta_{j} \rightarrow \beta_{j+1}\}, 1 \le j \le l-1$ be the pair of {\it partially overlapping chains of transitive dependencies} with all the requirements as in ~\defref{transChainsOverlap}. Assume that $\mathcal{R} = \{R_1, R_2, \ldots, R_m\}$ is a $2NF$ decomposition of $\Omega$. We may assume, without loss of generality\footnote{Otherwise - we may assume the rest of the database has been normalized to be precisely in 2NF, and the remaining part of the database to be normalized is that of $\left\{ \alpha_1 , \beta_1 \right\} ^{+}$.}, that $K = \{\alpha_1, \beta_1\}$. 

The proof is by contradiction. Assume that $\mathcal{R}$ is precisely in $2NF$. Without loss of generality, let $\gamma \in R_m$. Denote by $K_m$ the key of $R_m$. Consider the following two cases:
\begin{itemize}
\setlength\itemindent{-10pt}
\item $K_m = \alpha_1 \beta_1$: % first half
Since $k \geq 2$, there are two subcases to be considered:
\begin{itemize}
{\setlength\itemindent{-30pt}
\item $\exists i \ge 2 \, $ s.t. $\alpha_i \in R_m$:
%(resp. $\exists j \ge 2 \,$ s.t. $\beta_j \in R_m$)
That is, $\alpha_i$ 
%(resp. $\beta_j$) 
is partially dependent in the key, in contradiction to that $R_m$ is in $2NF$.
\item $\forall i \ge 2 : \alpha_i \notin R_m$ 
%and $\forall j \ge 2 : \beta_j \notin R_m$
. Consider the attributes $\alpha_{k^{*}}, \beta_{l^{*}}$. Then the functional dependency $\{\alpha_{k^{*}}, \beta_{l^{*}}\} \rightarrow \gamma$ has been lost.
}\end{itemize}
\item $K_m \neq \alpha_1 \beta_1$: Then, % second half
since $\alpha_1 \beta_1 \notin R_m$, the functional dependencies $\alpha_1 \beta_1 \rightarrow \alpha_{k^{*}} \beta_{l^{*}}$ and $\alpha_{k^{*}} \beta_{l^{*}} \rightarrow \gamma$ belong to distinct $R_{i}'s$, which means that a decomposition based on transitivity took place, contradicting that the system is precisely in $2NF$.
\begin{flushright}
$\Box$
\end{flushright}
% \begin{itemize}
%\item $\alpha_k \beta_l \in R_m$: $\exists i \ge 1 \, s.t. \, \alpha_i \notin R_m, \alpha_{i+1} \in R_m$ (resp. $\exists j \ge 1$ s.t. $\beta_j \notin R_m, \beta_{j+1} \in R_m$):
%That is,  $\{\alpha_{i} \rightarrow \alpha_{i+1}\}$ (resp. $\{\beta_{j} \rightarrow \beta_{j+1}\}$) has been lost. \item $\forall i \ge 2 : \alpha_i \notin R_m$ and $\forall j \ge 2 : \beta_j \notin R_m$: Consider the attributes $\alpha_k, \beta_l$. Then the functional dependency $\{\alpha_k, \beta_l\} \rightarrow \gamma$ has been lost.
%\end{itemize}
\end{itemize}
% having, for these $k,l \ge 2$, a `meeting point' in which $\{\alpha_{k}\beta_{l} \rightarrow \gamma\}$, where all $\alpha_1,\ldots,\alpha_{k}, \beta_1,\ldots,\beta_{l}, \gamma$ are disjoint and, for any $\delta \subset \alpha_{k}\beta_{l}$, it holds that $\delta \rightarrow \gamma \notin F^{+}$.

\end{proof}

We note that the criterion described in \thmref{no2NF} is not necessary. We demonstrate it with the following example.
\begin{example}\labelexm{notiff}
{\rm \fontfamily{cmss} \selectfont Consider the following $\Omega = \{\underline{A_1, A_2}, A_3, A_4\}$ with $$F = \{A_{1}A_{2} \rightarrow A_{3}, A_{3} \rightarrow A_{4}, A_{1}\rightarrow A_{4} \}.$$
The two `natural' decompositions are:
\begin{itemize}
    \item $\mathcal{R}_{a}=\left\{R_{1}\right\},\qquad R_{1} = \left\{\underline{A_1,A_2}, A_3, A_4\}\right\}.$ The attribute $A_4$ depends on $A_1\subset \left\{A_1,A_2\right\}$, therefore the decomposition is not in $2NF$. 
    \item $\mathcal{R}_{b}=\left\{R_{1}, R_{2}, R_{3}\},\qquad R_{1} = \{\underline{A_1,A_2}, A_3\}, R_{2} = \{\underline{A_3}, A_4\}, R_{3} = \{\underline{A_1}, A_4\right\}.$ However, here, the table $R_3$ was decomposed from (the original) $R_1$ due to transitivity. 
\end{itemize}
Any other decomposition is either lossy or does not preserve dependencies. Thus, for this $F$, one cannot be precisely in $2NF$.}
\end{example}

%%%%%%%%%%%%%%%%%%%%%%%%%%%%%%%%%%%%%%%%%%%%%%%%
%%%%%%%%%%%%%%%%% Acknoeledgement %%%%%%%%%%%%%%%%% 
%%%%%%%%%%%%%%%%%%%%%%%%%%%%%%%%%%%%%%%%%%%%%%%%
\subsubsection*{\fontfamily{cmss} \selectfont Acknowledgement}~\labelsubsec{acknowledge}
The authors would like to thank E. Gudess of BGU for fruitful discussions of the subject.

%%%%%%%%%%%%%%%%%%%%%%%%%%%%%%%%%%%%%%%%%%%%%%%%
%%%%%%%%%%%%%%%% Bibliogarphy %%%%%%%%%%%%%%%%%%
%%%%%%%%%%%%%%%%%%%%%%%%%%%%%%%%%%%%%%%%%%%%%%%%
% the bibliography is included in the DbsThesis.bib file
%\newpage
%\clearpage
\renewcommand*{\bibfont}{\small}
\begingroup
\phantomsection\printbibliography[heading=bibintoc]
\endgroup

\end {document}